# Enhanced Dirac node separation in strained $Cd_3As_2$ topological semimetal

G. Krizman,[1] J. Bermejo Ortiz,[2] M. Goyal,[3] A. Lygo,[3] J. Wang,[4] Z. Zhang,[5] B.A. Assaf,[4] S. Stemmer,[3] L.A. de Vaulchier,[2] Y. Guldner[2]

[1] *Institut für Halbleiter und Festkörperphysik, Johannes Kepler Universität, Altenbergerstrasse 69, 4040 Linz, Austria*

[2] *Laboratoire de Physique de l'Ecole normale supérieure, ENS, Université PSL, CNRS, Sorbonne Université, 24 rue Lhomond 75005 Paris, France*

[3] *Materials Department, University of California, Santa Barbara, CA 93106, USA*

[4] *Department of Physics and Astronomy, University of Notre Dame, Notre Dame, IN 46556, USA*

[5] *Advanced Photon Source, Argonne National Laboratory, Lemont, IL 60439, USA*

**In topological semimetals, nodes appear at symmetry points in the Brillouin zone as a result of band inversion, and yield quasi-relativistic massless fermions at low energies. $Cd_3As_2$ is a three-dimensional topological semimetal that hosts two Dirac cones responsible for a variety of quantum phenomena. In this work, we demonstrate the strain tuning of the Dirac nodes of $Cd_3As_2$ through a combination of magnetooptical infrared spectroscopy and high-resolution X-ray diffraction studies performed on epitaxial films. In these thin films, we observe a giant enhancement of the node separation in momentum space by close to a factor of 4. A combination of experimental measurements and theoretical modelling allows relate the origin of this enhancement to a strengthening of the topological band inversion driven by lattice strain. Our results demonstrate how strain can be used as a knob to tune the topological properties of semimetals and to potentially enhance their performance and response for various applications.**

## I. INTRODUCTION

Semimetals are materials that host a band overlap in their electronic structure while retaining a low density of state at the Fermi energy. With the rise of topological phases of matter, new semimetals were discovered, with peculiar band crossings. [1] In three-dimensions, such band crossings can occur at a finite number of points in the Brillouin zone, yielding what is referred to Dirac or Weyl fermions. [2] [3] [4] The main distinction between the two classes lies in the fact that Weyl fermions are intrinsically helical and exhibit spin-momentum locking. They generally occur in pairs of opposite helicity. Their properties have attracted tremendous attention. They are thought to host a chiral anomaly manifesting as a negative magnetoresistance [5] [4] [6] and have also been found to yield strong non-linear optical effects, [7] [8] thermoelectric effects, [9] efficient spin-charge conversion [10] [11] and spin transport. [12]

$Cd_3As_2$ was among the first discovered topological semimetals [13] [14]. This high quality stoichiometric material possesses two Dirac nodes. Thus, it is a first-choice system to study Weyl node physics in its simplest form. It crystallizes in a tetragonal structure that remains nearly cubic. Its electronic structure is well described by a modified Kane model similar to that of strained III-V and II-VI semiconductors. [14] [15] [16] In this model, two s-like (S) and six p-like bands (light hole (LH), heavy hole (HH), and split-off (SO)) interact near the Γ–point. In $Cd_3As_2$, a band inversion brings the s-like band below 4 of the p-like bands yielding a semimetal similar to HgTe. [17] [18] However, the natural tetragonality of

Cd$_3$As$_2$, i.e. the lattice elongation along the z//[001] direction, lifts the degeneracy of LH and HH causing them to cross at finite k$_z$. This effect yields two Dirac nodes in Cd$_3$As$_2$ with strongly electron-hole asymmetric dispersions, and a node separation governed by the splitting of the p-like bands and the tetragonal distortion (Fig. 1(a)). This asymmetric picture of the band structure of Cd$_3$As$_2$ is reproduced in magnetooptical measurements by independent groups, [19] [20] density-functional theory calculations, [21] [22] [13] scanning tunneling microscopy, [23] and early angle resolved photoemission measurements. [3] The fact that the origin of the node separation in Cd$_3$As$_2$ can be traced back to specific band and structural origins [20] [24] [25] makes it a prototypical topological semimetal ideal for band engineering.

Here, we demonstrate such a band engineering by lattice strain in Cd$_3$As$_2$ thin films grown by molecular beam epitaxy on Al$_{1-x}$In$_x$Sb buffer layers in (001) GaSb substrates. Using magnetooptical infrared spectroscopy measurements and $\bm{k}.\bm{p}$ modelling, we reveal a large enhancement of the Dirac nodes separation driven by this lattice strain (Fig. 1(b)). This translates into an enhancement of the Weyl node separation in Cd$_3$As$_2$. Our findings are compared to previous measurements on unstrained films for which the Dirac node separation is found to be much smaller (Fig. 1(a)). [20] They are corroborated by temperature dependent X-ray diffraction and magnetooptical measurements that can only be consistently explained by the picture and band ordering shown in Fig. 1(b).

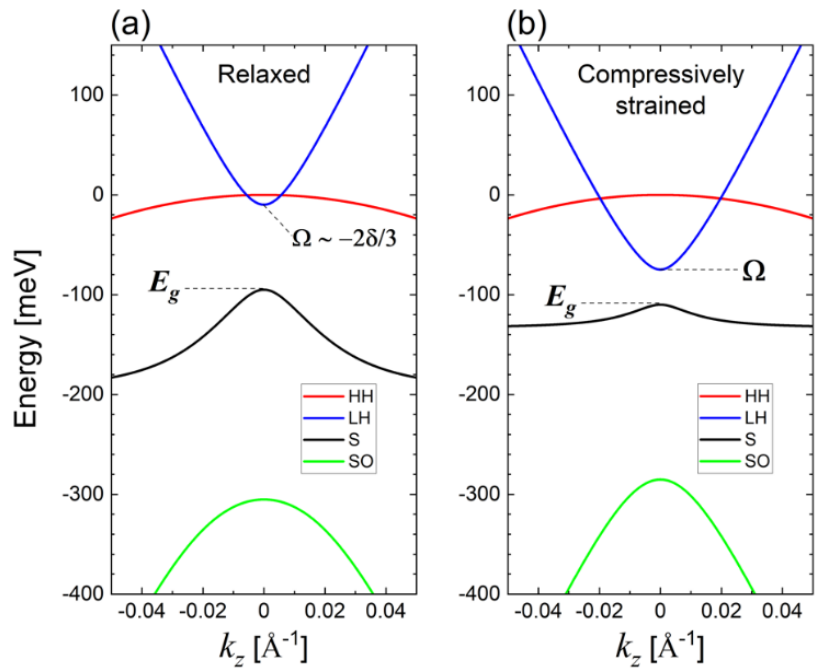

**FIG 1.** Band dispersion of **(a)** pristine ($c/2a \approx 1.005$) and **(b)** compressively strained Cd$_3$As$_2$ in the (001) plane achieved in this work ($c/2a \approx 1.023$). The bands are labeled in the standard notation of the Kane model, LH: light holes, HH: heavy holes, S: s-like band, SO: split-off band. Energy is taken to be 0 at the HH band edge. $\Omega$ represents the position of the LH band edge, and quantifies the LH-HH splitting, E$_g$ is the energy gap between HH and S.

## II. STRAIN CHARACTERIZATION

(001)-oriented Cd$_3$As$_2$ thin films (200 nm) are grown using molecular beam epitaxy on III-V substrates following what is shown in the following work. [26] We focus on two (001) oriented layers grown on a 2μm-thick Al$_{1-x}$In$_x$Sb buffer layer with x=0.5 as discussed in ref. [27], which are 150 and 200nm thick.

X-ray diffraction is carried out at the Advanced Photon Source at Argonne National Lab using beamline 33-ID-D. XRD patterns obtained at different temperatures are shown in Fig. 2(a). The Cd$_3$As$_2$ (00 16) Bragg peak can be seen at lower diffraction angle than the (004) peaks of the buffer layer and the GaSb substrate. We first analyze the lattice parameters of Cd$_3$As$_2$ at high temperature. At 260K, we find the out-of-plane parameter $c = 25.660 \pm 0.005$Å. A reciprocal space map obtained at 260K near the GaSb (224) peak is shown in Fig. 2(c) and allows us to determine the in-plane lattice parameter $a = 12.540 \pm 0.003$Å from the (44 16) peak of Cd$_3$As$_2$. From Fig. 2(b), we find that the Cd$_3$As$_2$ epilayer remains coherently strained to the underlying buffer layer that has an in-plane lattice parameter $a_{buffer} = a/2 = 6.27$Å. Comparing $c/2a$ to what is found for bulk single crystals near room temperature and at 100K from previous works [28] ($a_{bulk} = 12.67 + \pm0.01$ and $c_{bulk} = 25.48 \pm 0.02$) confirms that the unit cell is compressively strained in the plane, and elongated along the $c$-axis. The tetragonality $c/2a$ of the crystal is enhanced from 1.0055 (300K) [29] or 1.0064 (100K) [30] in (unstrained) bulk single crystals to more than 1.023 in our strained film. As detailed in the Appendix A, we evaluate the strain tensor as $\varepsilon_{xx} = \varepsilon_{yy} = -1.0 \pm 0.1\%$ and $\varepsilon_{zz} = +0.7 \pm 0.1\%$ when comparing our lattice parameters to those of pristine Cd$_3$As$_2$ from Ref. [28]. The reciprocal maps in Fig. 2(b,c) show that the high compressive biaxial strain is maintained in our 200nm-thick epilayer, as denoted by the sharpness of the peak attributed to the Cd$_3$As$_2$ lattice. Such a coherent compressive strain over a large thickness was previously reported in III-V heterostructures [31].

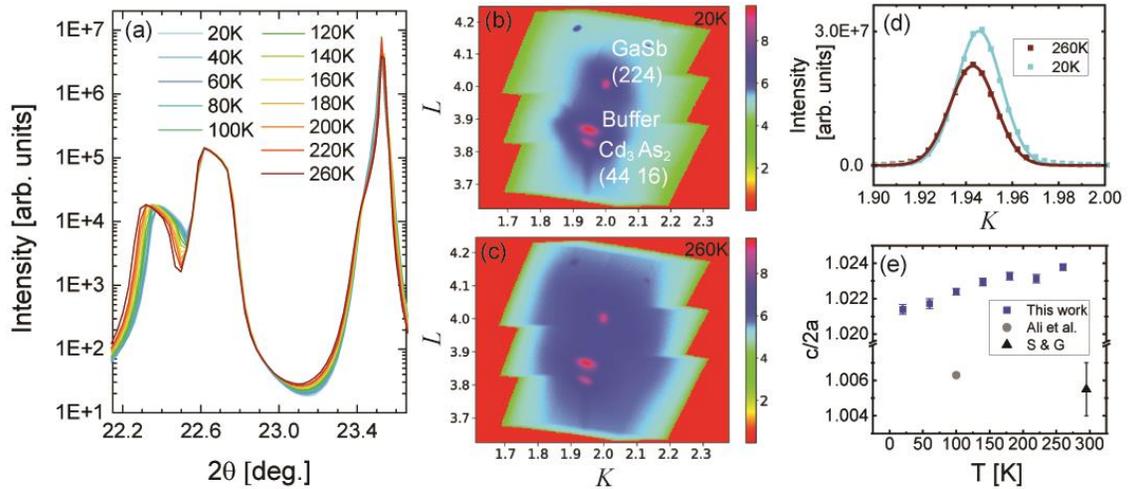

**FIG 2. (a)** Temperature dependent specular X-ray diffraction patterns taken using a beam energy of 20keV. **(b)** Reciprocal space maps obtained about the GaSb (224) peak at 20K and **(c)** at 260K. $L$ and $K$ are the Miller indices. **(d)** Integrated intensity of the Cd$_3$As$_2$ and buffer layer shown in the RMSs at two different temperatures versus K. **(e)** Experimental ratio $c/2a$ (blue squares) versus temperature compared to Ali et al. (ref [30] at 100K) and S&G (ref [29] at 300K).

The Cd$_3$As$_2$ peak shifts towards higher Bragg angles at the low temperature as seen in Fig. 2(a) indicating a reduction of $c$ (thermal contraction). To analyze the temperature dependence of $a$, reciprocal space maps (RSMs) were obtained down to 20K near the GaSb (224) peak (see Fig. 2(b,c)). Qualitatively, it is clear that the Cd$_3$As$_2$ (44 16) maintains its alignment with the buffer layer peak in the in-plane direction, indicating that the two have the same $a$-lattice constant, even at low temperature. The integrated intensity of the (44 16) sample and the buffer layer peaks as function of Miller index K are then determined by isolating a region of interest that excludes the GaSb peak. Curves of this integrated intensity versus K are shown for T=260K and T=20K in Fig. 2(d) and from which a slight reduction of the $a$ lattice constant can be extracted at each temperature. Visually, it is obvious that a lattice constant is slightly smaller at low temperature. Combining the lattice constants obtained from the Cd$_3$As$_2$ (00 16) and (44 16) peaks at various temperatures RSMs obtained at various temperature between 260K and 20K, we track the variation of the tetragonality given by the ratio $c/2a$ down to low temperatures of interest. $c/2a$ decreases as temperature drops, but remains much larger than what is found in bulk single crystals or relaxed layers. This means that, beyond the intrinsic distortion of Cd$_3$As$_2$ yielding $c/2a \sim 1.006$, [30] we have achieved a significant additional biaxial compressive strain resulting in $1.020 < c/2a < 1.024$. Furthermore, the slight reduction of the tetragonality as temperature decreases is attributed to the difference between the thermal expansion coefficients of the buffer and Cd$_3$As$_2$ layers. [32] [33] The lower thermal contraction of Al$_{1-x}$In$_x$Sb induces a small biaxial tensile stress on Cd$_3$As$_2$.

### III. STRAIN EFFECTS ON THE BAND STRUCTURE

We next study the electronic structure of this highly compressively strained Cd$_3$As$_2$ film using magnetooptical infrared spectroscopy measurements that are carried out at various temperatures. We use a Fourier Transform Infrared spectrometer coupled to a cryostat equipped with a 15T coil, as in our previous works. Detection is carried out using a composite Si bolometer and a HgCdTe detector. Magnetooptical spectra obtain at 4.2K in the mid-infrared are shown in Fig. 3(a) and those in the far-infrared in Fig. 3(b). Transition minima in Fig. 3(a,b) are due to inter-Landau level transitions, whose study versus magnetic field aims to extract the zero field band properties, mainly $E_g$ and Ω. The energy positions of the minima versus magnetic field is extracted and plotted in Fig. 3(c). To model these magnetooptical transitions, an 8-band Kane model is utilized [15] [19] and carefully detailed in the Appendix B1-2 and C. [15][17] [34] [35] The tetragonal distortion inherent to Cd$_3$As$_2$ introduces a crystal field splitting $\delta$ into the Kane model. [15,35] This parameter is directly proportional to the tetragonality $c/2a$ determined by XRD :

$$\delta = 3b\left(1 - \frac{c}{2a}\right) \qquad (1)$$

where $b < 0$ is a shear deformation potential, as it is defined in Ref [36]. It creates a direct relation between the crystalline and the electronic properties of the material. $\delta$ is the main parameter that contributes to the non-zero splitting of LH and HH, which we call here Ω (see Fig. 1). However, the spin-orbit parameter Δ, responsible for the splitting between LH and SO can also play a crucial role in Ω. It is because strain induces a coupling between the LH and SO bands (see Appendix B4) [37] [36,38]. The magnitude of the splitting responsible for the Dirac node creation is given by:

$$\Omega = -\frac{\Delta + \delta}{2} + \frac{1}{2}\sqrt{(\Delta - \delta)^2 + 4\frac{\delta\Delta}{3}} \tag{2}$$

In most of semiconductors [39,40], $\Delta$ is large enough (or $\delta$ is small enough ($\Delta \gg \delta$)) to neglect its influence on $\Omega$ and one finds $\Omega \approx -2\delta/3 = -2b(1 - c/2a)$, see Fig. 1(a). As we will see later, the temperature dependence of the magneto-optical oscillations determined in this work shows that the effect of $\Delta$ in Eq. (2) is non-negligible for highly strained samples, see Fig. 1(b).

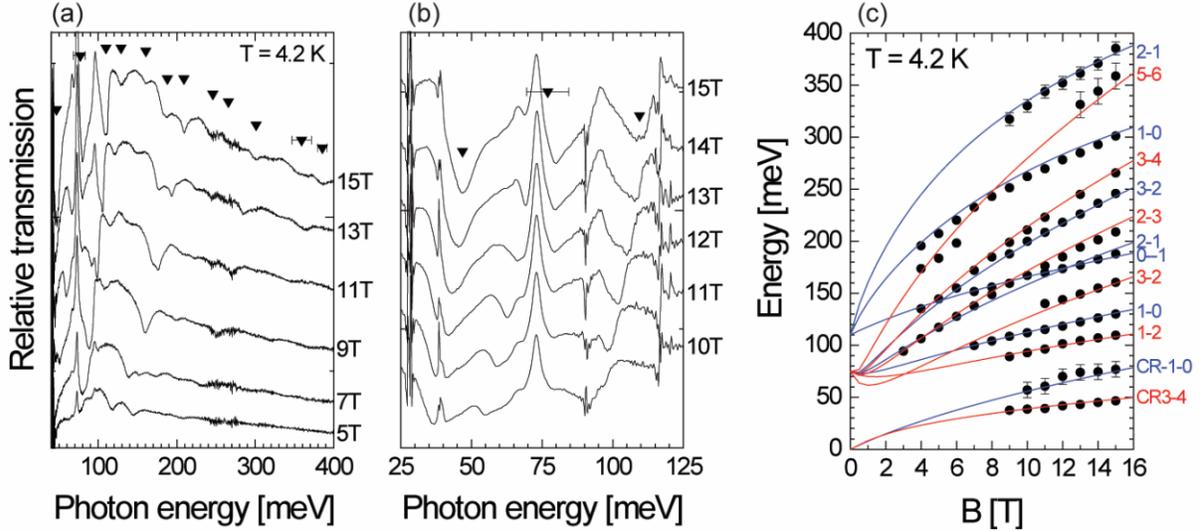

**FIG 3.** Magnetooptical spectra in mid-infrared **(a)** and far-infrared **(b)**. **(c)** Fan-chart showing data as the black points and calculated magnetooptical transitions in red and blue.

The experimental data allows us to extract with precision the band parameters. In particular, the positions of the bands are given by the extrapolation to zero field of the two series of magneto-optical oscillations (see Fig. 3(c)). At 4.2K we find $E_g = -110$ meV, implying that S lies below HH, and most importantly $\Omega = -75$ meV demonstrating the enhanced LH-HH splitting. The slope of the transitions at high energy (the bending of transitions "1-0" and "2-1" in blue with increasing magnetic field) allows an accurate determination of $\Delta = 220$ meV (independently from $\delta$) in excellent agreement with DFT calculations [13] [21]. The fit thus results in $\delta = 150$ meV following Eq. (2). This value is considerably enhanced compared to other works, performed on relaxed samples and single crystals, that give $\delta < 30$ meV [19,20,23,34]. Therefore, the corresponding Dirac node spacing found in this work is 0.04 Å$^{-1}$ (see Fig. 1(b)), which is four times greater than in relaxed layers or bulk crystals (see Fig. 1(a)). This is directly due to the much more pronounced tetragonality of the sample studied here, as it is shown by Eq. (1). A deformation potential $b = -2.1 \pm 0.5$ eV reconciles all these results and demonstrates the strain origin of the enhanced Dirac node separation. The determination of such deformation potential $b$ with respect to a cubic lattice allows for a direct comparison with other semiconductors. This parameter is universal and accounts for the effect of a (001) biaxial strain applied to any diamond and zinc-blende cubic lattices, i.e., it determines the strength of the p-type bands splitting under applied strain [36,38]. For diamond semiconductors, one can find in the literature $b = -2.1$ eV for Si, or $b = -2.5$ eV for Ge [41]. Zinc-blende semiconductors (GaAs, InAs, InSb, AlSb, InP, …) show $-2 < b < -1.3$ eV [41]. Therefore, by determining $b = -2.1$ eV, this work establishes Cd$_3$As$_2$ as a highly strain sensitive material.

## IV. TEMPERATURE DEPENDENCE OF Cd$_3$As$_2$ BAND STRUCTURE

Magneto-optical measurements have been performed up to $T = 200$ K. The analysis conducted in Sec. III for T=4 K is repeated for each temperature. This process results in optimal fits that are presented in Fig. 4. For each temperature, a great agreement is obtained with the experimental data, which allow an accurate determination of the band parameters: $E_g, \delta, \Delta$, and $P_\perp$ (or $v$), whose values are listed in Table I. The energy gap $E_g$ and the in-plane matrix elements $P_\perp$ are found to be almost temperature independent. The variations of $\delta(T)$ and $\Delta(T)$ can be unraveled by focusing on the first interband transition (labelled "1-2" in red in Fig. 3(c) and 4) as displayed in Fig. 5(a).

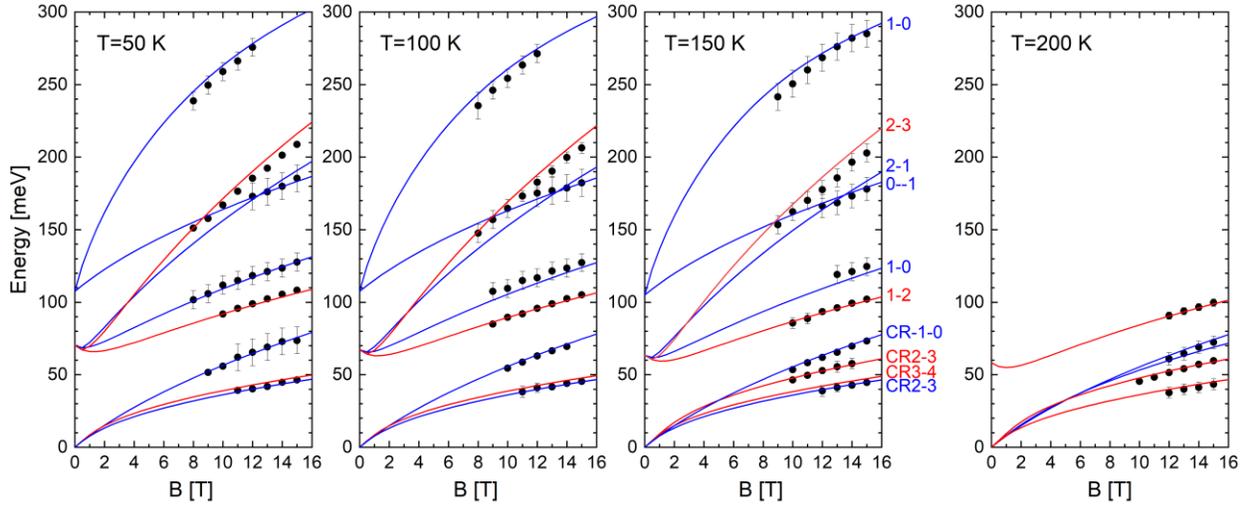

**FIG 4.** Magneto-optical measurements at different temperatures from 50 to 200 K. Dots represent the experimental data. Solid lines correspond to the calculated inter Landau level transitions using the fitting parameters listed in Table I.

The temperature dependence of the magneto-optical oscillations highlights the strong influence of the split-off band on $\Omega$. By increasing the temperature, the first interband transition undergoes a redshift as seen in Fig. 5(a), meaning that $|\Omega|$ is decreased as represented by the fits in Fig. 5(b). From the XRD measurements showing a slight increase of the tetragonality (see Fig. 2(e)), we know that $\delta$ increases by ~15 meV between 4 and 200K, as demonstrated by Eq. (1). This results in a slight increase of $|\Omega|$ with temperature, however, we unambiguously observe the opposite behavior, meaning that the evolution of $\Omega(T)$ is mainly governed by $\Delta(T)$, see Eq. (2). Overall, the measured evolution of $\Omega(T)$ can only be explained by a decreasing $\Delta$ when the temperature is increased, going from 220 meV to 120 meV at 4K and 200K, respectively. This evolution also confirms that the LH and HH (p-like) bands must lie above the S and SO bands.

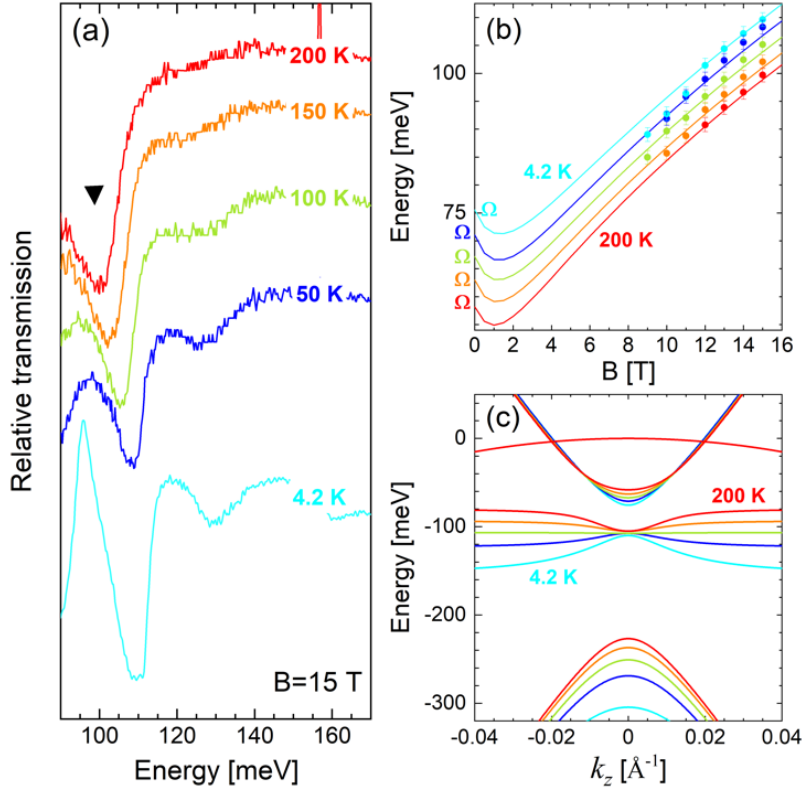

**FIG 5. (a)** Magneto-optical spectra at B=15 T for different temperatures, waving the attention on the first interband absorption (black arrow) that is redshifted when temperature is increased. **(b)** Fit of the first interband at each temperature, giving accurately $|\Omega|(T)$ from the extrapolation at zero field. **(c)** Band structure resulting from the magneto-optical fits at different temperature.

The fits result in the determination of the band structure plotted in Fig. 5(c). One can note the nearly temperature independent Dirac node spacing, due to the increased band edge mass when $\Omega(T)$ is decreased. Interestingly, at 100K, we find $\Delta \approx \delta \approx 160$ meV, meaning that the LH-SO mixing is unusually strong. Indeed, we determine that the SO component weights from 11% at 4K to 30% at 200K on the LH band edge (see Appendix B4). For temperature higher than 100K, the curvature of the S band is even reversed as seen in Fig. 5(c). This phenomenon appears when $\Delta < -3E_g/2$ and is an original consequence of (i) the small spin-orbit parameter found in this compound and (ii) its inverted band structure (see Appendix B3).

**Table I.** Band parameters leading to the best fit of the magneto-optical data in temperature.

| T [K] | $E_g$ [meV] | $\Delta$ [meV] | $\delta$ [meV] | $v$ [x10$^6$ m/s] |
|---|---|---|---|---|
| 4 | -110 ±2.5 | 220 ±10 | 150 ±5 | 0.97 ±0.01 |
| 50 | -107.5 ±2.5 | 180 ±10 | 155 | 0.97 ±0.01 |
| 100 | -107.5 ±2.5 | 160 ±10 | 158 | 0.96 ±0.01 |
| 150 | -105 ±2.5 | 140 ±10 | 162 | 0.95 ±0.01 |
| 200 | -105 ±5 | 120 ±20 | 165 | 0.95 ±0.01 |

## V. CONCLUSION

We have thus shown that the Dirac nodes in Cd$_3$As$_2$ can be tuned by strain. By understanding the intrinsic tetragonal distortion of Cd$_3$As$_2$ to be the cause of Dirac nodes creation, we have been able to enhance the separation of these Dirac nodes using compressive biaxial strain. XRD and magneto-optical spectroscopy have probed the giant strain-induced enhancement of this separation. Our work thus demonstrates that many properties of Cd$_3$As$_2$ that are driven by its topological band structure, can be largely enhanced by strain. Those may include non-linear optical phenomena [7] and spintronic conversion [12].

The temperature dependence of the Cd$_3$As$_2$ band structure that we elucidate, strongly reinforces our analysis and highlights the importance of the often-neglected strain-induced mixing of the LH and SO bands. This mixing is of interest as it drastically alters, among others, the spin properties ($g$-factor) and is therefore, important to consider for the conception of efficient quantum-dot based spin qubits when LH states are involved [16,42–45]. This unique property certainly requires further investigations in Cd$_3$As$_2$ films. [36]

## ACKNOWLEDGEMENTS

This work is supported by ANR-19-CE30-022-01. M.G. A.L. and S.S. are supported by Vannevar Bush Faculty Fellowship program by the U.S. Department of Defense for support (Grant No. N00014-16-1-2814). JW and BAA acknowledge support from NSF-DMR-1905277. This research used resources of the Advanced Photon Source, a U.S. Department of Energy (DOE) Office of Science user facility operated for the DOE Office of Science by Argonne National Laboratory under Contract No. DE-AC02-06CH11357.

## APPENDIX A: STRAIN CHARACTERIZATION

The strain is determined with respect to a bulk Cd$_3$As$_2$ which lattice parameters are measured in Ref. [28]. A pristine Cd$_3$As$_2$ crystallizes in a tetragonal lattice with $a_{bulk} = b_{bulk} = 12.67$ Å and $c_{bulk} = 25.48$ Å. The XRD measurements presented in the main text give, at high temperature, $a = b = 12.540 \pm 0.003$ Å and $c = 25.660 \pm 0.005$ Å. Therefore, the lattice mismatches between our strain epilayers and the bulk are calculated as:

$$\begin{cases} \varepsilon_\parallel = \dfrac{a - a_{bulk}}{a_{bulk}} = -0.010 \pm 0.001 \\ \varepsilon_\perp = \dfrac{c - c_{bulk}}{c_{bulk}} = +0.007 \pm 0.001 \end{cases}$$

The in-plane strain in our samples is thus compressive and around 1 %. From the determination of $\varepsilon_\parallel$ and $\varepsilon_\perp$, one can deduce the Poisson ratio of Cd$_3$As$_2$ under a (001) biaxial strain (at room temperature):

$$\lambda = 2\frac{C_{12}}{C_{11}} = \frac{\varepsilon_\perp}{\varepsilon_\parallel} = 0.7$$

## APPENDIX B: MODIFIED KANE MODEL

### 1. Band structure of tetragonal Cd₃As₂

The tetragonal lattice of Cd$_3$As$_2$ being nearly cubic, we model this system with the Kane Hamiltonian developed for cubic lattices, in which a biaxial strain has been implemented [15,34–36,38]:

$$H_k = \begin{pmatrix} E_g & P_\perp k_- & -P_\perp k_+ & 0 & 0 & 0 & 0 & P_\parallel k_z \\ P_\perp k_+ & -\frac{\hbar^2 k_z^2}{2\widetilde{m}} & 0 & 0 & 0 & 0 & 0 & 0 \\ -P_\perp k_- & 0 & -\frac{2\Delta}{3} & \frac{\sqrt{2}\Delta}{3} & 0 & 0 & 0 & 0 \\ 0 & 0 & \frac{\sqrt{2}\Delta}{3} & -\left(\delta+\frac{\Delta}{3}\right) & P_\parallel k_z & 0 & 0 & 0 \\ 0 & 0 & 0 & P_\parallel k_z & E_g & P_\perp k_+ & P_\perp k_- & 0 \\ 0 & 0 & 0 & 0 & P_\perp k_- & -\frac{\hbar^2 k_z^2}{2\widetilde{m}} & 0 & 0 \\ 0 & 0 & 0 & 0 & P_\perp k_+ & 0 & -\frac{2\Delta}{3} & \frac{\sqrt{2}\Delta}{3} \\ P_\parallel k_z & 0 & 0 & 0 & 0 & 0 & \frac{\sqrt{2}\Delta}{3} & -\left(\delta+\frac{\Delta}{3}\right) \end{pmatrix} \quad (B1)$$

with $k_\pm = (k_x \pm k_y)/\sqrt{2}$ and $k_z$ along the [001] direction. The matrix is given in the following basis: $i|S\downarrow\rangle$, $|(X-iY)\downarrow\rangle/\sqrt{2}$, $-|(X+iY)\downarrow\rangle/\sqrt{2}$, $|Z\uparrow\rangle$, $i|S\uparrow\rangle$, $|(X+iY)\uparrow\rangle/\sqrt{2}$, $|(X-iY)\uparrow\rangle/\sqrt{2}$, $|Z\downarrow\rangle$.

$E_g$ is the band gap separating the S band from the HH band. A negative energy gap means that the S band lie below the HH band. Because the spin-degenerate HH band has no $k_z$-interaction with the other bands considered in this model, its $k_z$-dispersion is flat in the Kane model. However, second order interactions with other remote bands at higher or lower energies induce a small negative parabolic curvature for the HH bands, given by the effective mass $\widetilde{m} > 0$. This work does not pretend to provide any determination for $\widetilde{m}$ as the magnetic field is applied along the z-direction and thus, only probe the in-plane motions of electrons. We arbitrary fix it to $0.40\ m_0$ all over the manuscript. $P_\perp$ and $P_\parallel$ are respectively the in-plane and out-of-plane Kane matrix elements. They are related to the electron velocity as follow $v = \sqrt{2/3}\ P_\perp/\hbar$ [20]. $\Delta$ is the spin-orbit parameter.

$\delta$ is the crystal field splitting that models the tetragonal distortion. In fact, it is the energy lifting of the $|Z\rangle$ orbital (due to the lattice elongation along $z$) compare to the other two p-type orbitals $|X\rangle$ and $|Y\rangle$. This distortion has two effects: (i) the LH band is shift away from the HH band and thus, under a compressive biaxial strain ($\delta > 0$), make the two Dirac nodes appear; (ii) the Hamiltonian is no longer diagonalizable in a good basis of Bloch functions, thus, the LH and SO bands are mixed [38].

At $k = 0$, Eq. (B1) gives the twofold-degenerate four band energies:

$$\begin{cases} E(S) = E_g \\ E(HH) = 0 \\ E(LH) = -\dfrac{\Delta + \delta}{2} + \dfrac{1}{2}\sqrt{(\Delta - \delta)^2 + 4\dfrac{\delta\Delta}{3}} \\ E(SO) = -\dfrac{\Delta + \delta}{2} - \dfrac{1}{2}\sqrt{(\Delta - \delta)^2 + 4\dfrac{\delta\Delta}{3}} \end{cases} \quad (B2)$$

## 2. Deviation from a cubic lattice

In this model, the tetragonal deformation is considered with respect to a cubic lattice of parameter $a_0$, and underlies in the parameters $E_g$ and $\delta$. They can be revealed explicitly by replacing

$$\begin{cases} E_g(\tilde{\varepsilon}_\perp) = E_g(\tilde{\varepsilon}_\perp = 0) + A(2 - \lambda)\tilde{\varepsilon}_\perp - b(1 + \lambda)\tilde{\varepsilon}_\perp \\ \delta = 3b(1 + \lambda)\tilde{\varepsilon}_\perp \end{cases} \quad (B3)$$

where $A$ and $b$ are the hydrostatic and shear deformation potentials [36,38], $\lambda$ depends on the elastic constants ($\lambda = 2\,C_{12}/C_{11}$ for the distortion along [001] of our case). Note that $\tilde{\varepsilon}_\perp$ corresponds to the in-plane lattice mismatch with respect to a cubic lattice. It is given by:

$$\tilde{\varepsilon}_\perp = \frac{2a - a_0}{a_0}$$

with $a_0 = \dfrac{c - 2a\lambda}{1 + \lambda}$, the lattice parameter considering a cubic crystal. Therefore, one can write:

$$\tilde{\varepsilon}_\perp = \frac{1 - \chi}{\lambda + \chi}$$

where we have defined the tetragonality as $\chi = c/2a$. Because its value remains close to 1, Eq. (B3) can be simplified into:

$$\begin{cases} E_g(\tilde{\varepsilon}_\perp) = E_g(\tilde{\varepsilon}_\perp = 0) + A(2 - \lambda)\tilde{\varepsilon}_\perp - b(1 - \chi) \\ \delta = 3b(1 - \chi) \end{cases} \quad (B4)$$

The value of $b$ can be accurately extracted from our XRD and magneto-optical measurements that give respectively $\chi$ and $\delta$. The hydrostatic deformation potential $A$ can be deduced from the Poisson ratio of Cd$_3$As$_2$ determined in Appendix A. The band edges versus tetragonality (deviation from cubic) are calculated using Eq. (B2) and Eq. (B4) and plotted in Fig. 6 for $\lambda = 0.7$.

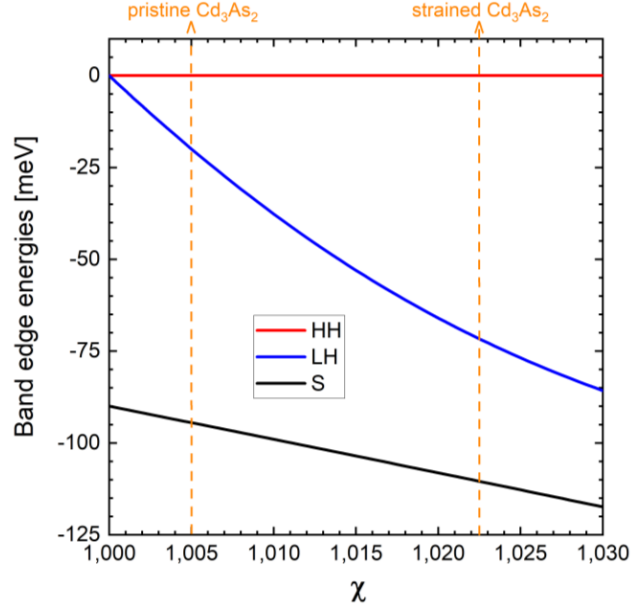

**FIG 6.** Band edges of Cd$_3$As$_2$ versus its tetragonality. $\lambda = 0.7$ is used for this figure, which gives $A = -1.6$ eV. The other parameters are determined as $b = -2.1$ eV, $E_g(\tilde{\varepsilon}_\perp = 0) = -90$ meV and $\Delta = 220$ meV.

### 3. $k_z$-dispersion

At $k_\perp = 0$, beside the HH band which is a parabolic band ($E = -\frac{\hbar^2 k_z^2}{2\tilde{m}}$), the three other dispersions are solutions of:

$$(E - E_g)[3E(\Delta + E) + \delta(2\Delta + 3E)] = (2\Delta + 3E)P_\parallel^2 k_z^2$$

which gives the behavior shown in Fig. 1. A Taylor expansion near $E_g$ gives the parabolic band edge mass of the S band:

$$\frac{\hbar^2}{2m^*} = P_\parallel^2 \frac{2\Delta + 3E_g}{3E_g(\Delta + E_g) + \delta(2\Delta + 3E_g)}$$

The S band reverses its curvature for $\Delta < -3E_g/2$, as observed in temperature (see Fig. 5(c)).

### 4. LH-SO mixing

The band energies obtained at $k = 0$ show that the mixing of LH and SO corresponds to an avoided-crossing between two levels of energy $-\Delta$ and $-\delta$ (see Eq. B2). The Bloch functions are mixed at $k = 0$ following [38]:

$$|SO\rangle = -\beta \left|\frac{3}{2}; \pm\frac{1}{2}\right\rangle + \alpha \left|\frac{1}{2}; \pm\frac{1}{2}\right\rangle$$

$$|LH\rangle = \alpha \left|\frac{3}{2}; \pm\frac{1}{2}\right\rangle + \beta \left|\frac{1}{2}; \pm\frac{1}{2}\right\rangle$$

With $|J; \pm m_J\rangle$ the Bloch functions basis for the cubic case, $|\alpha|^2 + |\beta|^2 = 1$, and

$$\begin{cases} \alpha = \dfrac{2|\delta|}{3\sqrt{n(n-p)}} \\ \beta = -\dfrac{\delta}{|\delta|}\sqrt{\dfrac{n-p}{2n}} \end{cases} \text{with} \quad n = \sqrt{p^2 + \dfrac{8}{9}\delta^2} \quad \text{and} \quad p = \Delta - \dfrac{\delta}{3}$$

## 5. Modelling an anisotropic in-plane strain

We consider here the effects of an anisotropic in-plane strain on Cd$_3$As$_2$ band structure, i.e, $\varepsilon_{xx} \neq \varepsilon_{yy}$. This situation is not realized in the present experimental work, but worth being theoretically addressed as it is experimentally feasible [46]. Bir and Pikus have determined that such a strain induces a $k$-independent interaction $M$ between p-type bands [36]. Therefore, the HH and LH band crossing is lifted by an anisotropic in-plane strain. Figure 7 shows the vanishing Dirac nodes under this crystal deformation. These dispersions have been calculated using Bir and Pikus strain Hamiltonian [36,47], which gives an interacting gap $2M$ that writes:

$$2M \sim -\sqrt{3}b(\varepsilon_{xx} - \varepsilon_{yy})$$

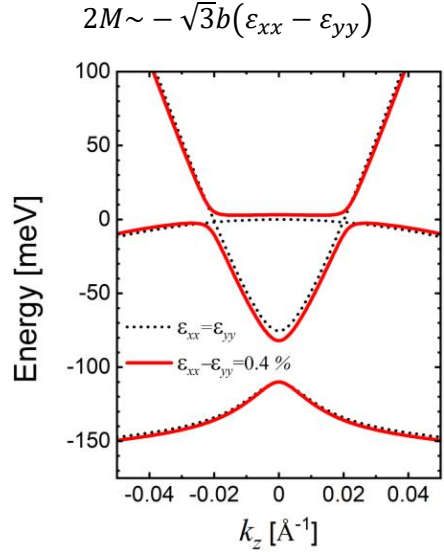

**FIG 7.** Calculated dispersions of Cd$_3$As$_2$ using parameters listed in Table I at T=4 K. In black dots are the dispersions under a homogeneous in-plane strain, while in red lines are the dispersions considering a huge anisotropic in-plane strain.

## APPENDIX C: LANDAU LEVELS AND MAGNETO-OPTICAL OSCILLATOR STRENGTH

Under an applied magnetic field along [001]//z, we perform the following Peierls substitution in the Hamiltonian (B1): $k_+ = \xi a^\dagger$ and $k_- = \xi a$, with $\xi = eB/\hbar$ the square magnetic length and $a$ and $a^\dagger$ the usual ladder operators. The Hamiltonian (B1) is then projected in a harmonic oscillator functions basis and gives:

$$H_B = \begin{pmatrix} E_g & P_\perp \xi \sqrt{n+1} & -P_\perp \xi \sqrt{n} & 0 & 0 & 0 & 0 & P_\parallel k_z \\ P_\perp \xi \sqrt{n+1} & -\dfrac{\hbar^2 k_z^2}{2\widetilde{m}} & 0 & 0 & 0 & 0 & 0 & 0 \\ -P_\perp \xi \sqrt{n} & 0 & -\dfrac{2\Delta}{3} & \dfrac{\sqrt{2}\Delta}{3} & 0 & 0 & 0 & 0 \\ 0 & 0 & \dfrac{\sqrt{2}\Delta}{3} & -\left(\delta+\dfrac{\Delta}{3}\right) & P_\parallel k_z & 0 & 0 & 0 \\ 0 & 0 & 0 & P_\parallel k_z & E_g & P_\perp \xi \sqrt{n} & P_\perp \xi \sqrt{n+1} & 0 \\ 0 & 0 & 0 & 0 & P_\perp \xi \sqrt{n} & -\dfrac{\hbar^2 k_z^2}{2\widetilde{m}} & 0 & 0 \\ 0 & 0 & 0 & 0 & P_\perp \xi \sqrt{n+1} & 0 & -\dfrac{2\Delta}{3} & \dfrac{\sqrt{2}\Delta}{3} \\ P_\parallel k_z & 0 & 0 & 0 & 0 & 0 & \dfrac{\sqrt{2}\Delta}{3} & -\left(\delta+\dfrac{\Delta}{3}\right) \end{pmatrix}$$

Where $n = -1, 0, 1, ...$ The magneto-optical transitions occur at $k_z = 0$, where the joint density of states is maximal. The Landau levels are calculated by diagonalizing $H_B(k_z = 0)$. The latter can be decoupled in two 4x4 blocks, whose eigenvalues correspond to the two spin components plotted in red and blue in the main text. The Landau levels corresponding to the fit performed at T=4 K (see Fig. 3(b) of the main text) are given in Fig. 8.

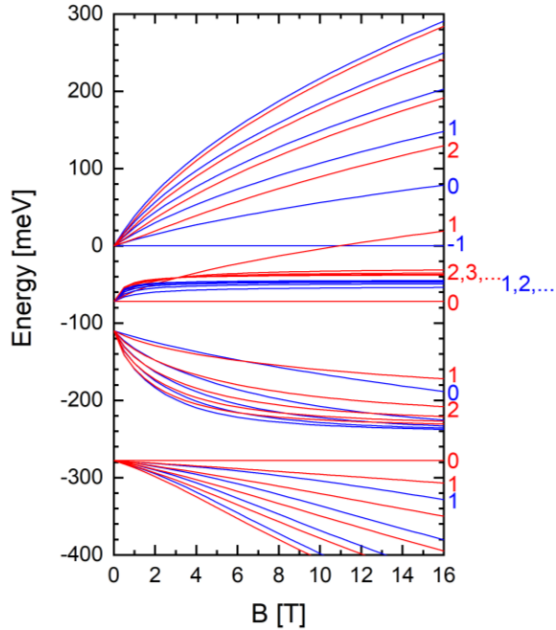

**FIG 8.** Calculated Landau levels of $Cd_3As_2$ giving the best fit of the magneto-optical data at T=4 K. Spins are in color and some Landau level indices are written at the right of the figure.

By writing the initial and final Landau levels involved in a magneto-optical transitions as the following:

$$|i\rangle = \begin{pmatrix} \alpha_{1,n}|n\rangle \\ \alpha_{2,n}|n+1\rangle \\ \alpha_{3,n}|n-1\rangle \\ \alpha_{4,n}|n-1\rangle \\ \alpha_{5,n}|n\rangle \\ \alpha_{6,n}|n-1\rangle \\ \alpha_{7,n}|n+1\rangle \\ \alpha_{8,n}|n+1\rangle \end{pmatrix} \quad and \quad |f\rangle = \begin{pmatrix} \beta_{1,n}|m\rangle \\ \beta_{2,n}|m+1\rangle \\ \beta_{3,n}|m-1\rangle \\ \beta_{4,n}|m-1\rangle \\ \beta_{5,n}|m\rangle \\ \beta_{6,n}|m-1\rangle \\ \beta_{7,n}|m+1\rangle \\ \beta_{8,n}|m+1\rangle \end{pmatrix}$$

The oscillator strengths of the transitions are proportional $|\langle f|d_\pm|i\rangle|^2$, where $d_\pm$ is the dipole operators of the $\sigma^+$ and $\sigma^-$ light polarization used in our case (Faraday geometry). These operators write:

$$d_\pm = \epsilon_\pm . v = \frac{1}{\hbar} \sum_{j=x,y,z} \epsilon_j \frac{\partial H_k}{\partial k_j}$$

With $\epsilon_\pm = (1/\sqrt{2}; \pm i/\sqrt{2}; 0)$ for the $\sigma^+$ and $\sigma^-$ polarization. Calculations give simply:

$$|\langle f|d_+|i\rangle|^2 = \frac{2P_\perp^2}{\hbar} |\beta^*_{1,n+1}\alpha_{2,n} - \beta^*_{3,n+1}\alpha_{1,n} + \beta^*_{5,n+1}\alpha_{7,n} + \beta^*_{6,n+1}\alpha_{5,n}|^2$$

$$|\langle f|d_-|i\rangle|^2 = \frac{2P_\perp^2}{\hbar} |-\beta^*_{1,n-1}\alpha_{3,n} - \beta^*_{2,n-1}\alpha_{1,n} + \beta^*_{5,n-1}\alpha_{6,n} + \beta^*_{7,n-1}\alpha_{5,n}|^2$$

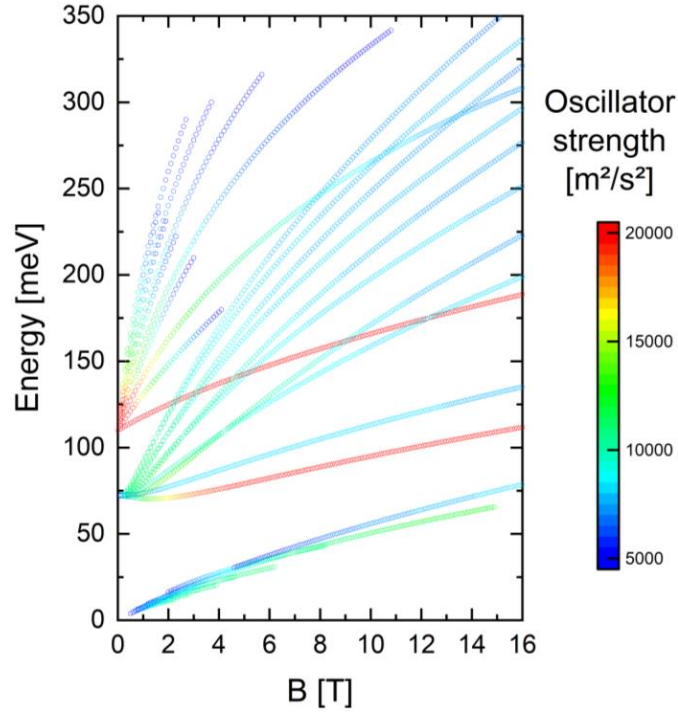

**FIG 9.** Calculated magneto-optical transitions between the Landau levels shown in Fig. 8. The corresponding oscillator strengths of the transitions are given using a color scale, in units of x$10^8$ m²/s². Only the transitions having a matrix element above 5000x$10^8$ m²/s² are considered.

The oscillator strengths of the most probable magneto-optical transitions have been calculated and are shown in Fig. 9 using a color scale. They greatly confirm our analysis presented in Fig. 3 of the main text, showing the most probable transitions. From the matrix elements, the selection rules can be deduced and lead to transitions between Landau levels of identical spin and indices that differ by one ($n \rightarrow n \pm 1$).